\documentclass[sigchi, authorversion, nonacm]{acmart}

\usepackage{booktabs} 
\usepackage{algpseudocode} 
\usepackage{algorithmicx }
\usepackage[toc,page]{appendix}
\usepackage{pdfpages}

\setcopyright{rightsretained}

\begin{document}
\title[Building Proactive Voice Assistants: When and How (not) to Interact]{Building Proactive Voice Assistants:\\When and How (not) to Interact}


\author{O. Miksik$^1$*, I. Munasinghe$^1$*, J. Asensio-Cubero$^1$*, S. Reddy Bethi$^1$*, S-T. Huang$^1$, S. Zylfo$^1$,\\
X. Liu$^1$, T. Nica$^1$, A. Mitrocsak$^1$,  S. Mezza$^1$, R. Beard$^1$, R. Shi$^1$, R. Ng$^1$, P. Mediano$^1$,\\
  Z. Fountas$^1$, S-H. Lee$^1$, J. Medvesek$^1$, H. Zhuang$^1$, Y. Rogers$^2$*, P. Swietojanski$^1$*}
\authornoteshared{*Equal contribution. Copyright 2020 by the authors.}
\affiliation{%
  \institution{$^1$Work done at Emotech Labs, U.K., $^2$ UCL Interaction Centre, University College London, U.K.}
}

\renewcommand{\shortauthors}{Miksik et al.}


\newcommand{\eq}{\textit{Eq.}}
\newcommand{\eg}{\textit{e.g.~}}
\newcommand{\ie}{\textit{i.e.~}}
\newcommand{\etal}{\textit{et al.}}

\newcommand\ignore[1]{}


\newcommand{\fig}{Fig.}
\newcommand{\figref}[1]{Fig.~\ref{#1}}

\newcommand{\cf}{\textit{cf.~}}
\newcommand{\etc}{\textit{etc.~}}
\newcommand*\diff{\mathop{}\!\mathrm{d}}
\newcommand{\ds}{\diff \s}
\newcommand{\dx}{\diff x}
\newcommand{\dy}{\diff y}

\newcommand{\transpose}{^\mathrm{T}}
\newcommand{\inv}{^{-1}}

\def \spanof{\mathop{\rm span}\nolimits}
\def \clip{\mathrel{\rm clip}}
\def \det{\mathop{\rm det}\nolimits}
\def \dim{\mathop{\rm dim}\nolimits}
\def \op{\mathrel{\rm op}}
\def \outcode{\mathop{\rm outcode}\nolimits}
\def \pixel{\mathop{\rm pixel}\nolimits}
\def \rank{\mathop{\rm rank}\nolimits}
\def \round{\mathrel{\rm round}}
\def \sgn{\mathop{\rm sgn}\nolimits}
\def \sinc{\mathrel{\rm sinc}}
\def \spur{\mathop{\rm spur}\nolimits}
\def \stencil{\mathop{\rm stencil}\nolimits}
\def \supp{\mathop{\rm supp}\nolimits}
\def \weci{\mathop{\rm wec}\nolimits}
\def \wec{\mathop{\rm WEC}\nolimits}
\def \zbuff{\mathop{\rm zBuffer}\nolimits}
\def \ease{\mathop{\rm ease}\nolimits}
\def \path{\bp C}
\def \profile{\bp S}


\newcommand{\bfa}{{\mathbf{a}}}
\newcommand{\bfb}{{\mathbf{b}}}
\newcommand{\bfc}{{\mathbf{c}}}
\newcommand{\bfd}{{\mathbf{d}}}
\newcommand{\bfe}{{\mathbf{e}}}
\newcommand{\bff}{{\mathbf{f}}}
\newcommand{\bfg}{{\mathbf{g}}}
\newcommand{\bfh}{{\mathbf{h}}}
\newcommand{\bfi}{{\mathbf{i}}}
\newcommand{\bfj}{{\mathbf{j}}}
\newcommand{\bfk}{{\mathbf{k}}}
\newcommand{\bfl}{{\mathbf{l}}}
\newcommand{\bfm}{{\mathbf{m}}}
\newcommand{\bfn}{{\mathbf{n}}}
\newcommand{\bfo}{{\mathbf{o}}}
\newcommand{\bfp}{{\mathbf{p}}}
\newcommand{\bfq}{{\mathbf{q}}}
\newcommand{\bfr}{{\mathbf{r}}}
\newcommand{\bfs}{{\mathbf{s}}}
\newcommand{\bft}{{\mathbf{t}}}
\newcommand{\bfu}{{\mathbf{u}}}
\newcommand{\bfv}{{\mathbf{v}}}
\newcommand{\bfw}{{\mathbf{w}}}
\newcommand{\bfx}{{\mathbf{x}}}
\newcommand{\bfy}{{\mathbf{y}}}
\newcommand{\bfz}{{\mathbf{z}}}
\newcommand{\bfell}{{\bolsymbol{ell}}}


\newcommand{\bfA}{{\mathbf{A}}}
\newcommand{\bfB}{{\mathbf{B}}}
\newcommand{\bfC}{{\mathbf{C}}}
\newcommand{\bfD}{{\mathbf{D}}}
\newcommand{\bfE}{{\mathbf{E}}}
\newcommand{\bfF}{{\mathbf{F}}}
\newcommand{\bfG}{{\mathbf{G}}}
\newcommand{\bfH}{{\mathbf{H}}}
\newcommand{\bfI}{{\mathbf{I}}}
\newcommand{\bfJ}{{\mathbf{J}}}
\newcommand{\bfK}{{\mathbf{K}}}
\newcommand{\bfL}{{\mathbf{L}}}
\newcommand{\bfM}{{\mathbf{M}}}
\newcommand{\bfN}{{\mathbf{N}}}
\newcommand{\bfO}{{\mathbf{O}}}
\newcommand{\bfP}{{\mathbf{P}}}
\newcommand{\bfQ}{{\mathbf{Q}}}
\newcommand{\bfR}{{\mathbf{R}}}
\newcommand{\bfS}{{\mathbf{S}}}
\newcommand{\bfT}{{\mathbf{T}}}
\newcommand{\bfU}{{\mathbf{U}}}
\newcommand{\bfV}{{\mathbf{V}}}
\newcommand{\bfW}{{\mathbf{W}}}
\newcommand{\bfX}{{\mathbf{X}}}
\newcommand{\bfY}{{\mathbf{Y}}}
\newcommand{\bfZ}{{\mathbf{Z}}}


\providecommand{\BA}{{\boldsymbol{A}}}
\providecommand{\BB}{{\boldsymbol{B}}}
\providecommand{\BC}{{\boldsymbol{C}}}
\providecommand{\BD}{{\boldsymbol{D}}}
\providecommand{\BE}{{\boldsymbol{E}}}
\providecommand{\BF}{{\boldsymbol{F}}}
\providecommand{\BG}{{\boldsymbol{G}}}
\providecommand{\BH}{{\boldsymbol{H}}}
\providecommand{\BI}{{\boldsymbol{I}}}
\providecommand{\BJ}{{\boldsymbol{J}}}
\providecommand{\BK}{{\boldsymbol{K}}}
\providecommand{\BL}{{\boldsymbol{L}}}
\providecommand{\BM}{{\boldsymbol{M}}}
\providecommand{\BN}{{\boldsymbol{N}}}
\providecommand{\BO}{{\boldsymbol{O}}}
\providecommand{\BP}{{\boldsymbol{P}}}
\providecommand{\BQ}{{\boldsymbol{Q}}}
\providecommand{\BR}{{\boldsymbol{R}}}
\providecommand{\BS}{{\boldsymbol{S}}}
\providecommand{\BT}{{\boldsymbol{T}}}
\providecommand{\BU}{{\boldsymbol{U}}}
\providecommand{\BV}{{\boldsymbol{V}}}
\providecommand{\BW}{{\boldsymbol{W}}}
\providecommand{\BX}{{\boldsymbol{X}}}
\providecommand{\BY}{{\boldsymbol{Y}}}
\providecommand{\BZ}{{\boldsymbol{Z}}}


\newcommand{\BGamma}{\mathbf{\Gamma}}
\newcommand{\bdelta}{\mathbf{\Delta}}
\newcommand{\btheta}{\mathbf{\Theta}}
\newcommand{\BLambda}{\mathbf{\Lambda}}
\newcommand{\BXi}{\mathbf{\Xi}}
\newcommand{\BPi}{\mathbf{\Pi}}
\newcommand{\BPsi}{\mathbf{\Psi}}
\newcommand{\BSigma}{\mathbf{\Sigma}}
\newcommand{\BUpsilon}{\mathbf{\Upsilon}}
\newcommand{\bphi}{\mathbf{\Phi}}
\newcommand{\BOmega}{\mathbf{\Omega}}


\newcommand{\nablabf}{\boldsymbol{\nabla}}
\newcommand{\alphabf}{\boldsymbol{\alpha}}
\newcommand{\betabf}{\boldsymbol{\beta}}
\newcommand{\gammabf}{\boldsymbol{\gamma}}
\newcommand{\deltabf}{\boldsymbol{\delta}}
\newcommand{\epsilonbf}{\boldsymbol{\epsilon}}
\newcommand{\varepsilonbf}{\boldsymbol{\varepsilon}}
\newcommand{\zetabf}{\boldsymbol{\zeta}}
\newcommand{\etabf}{\boldsymbol{\eta}}
\newcommand{\thetabf}{\boldsymbol{\theta}}
\newcommand{\varthetabf}{\boldsymbol{\vartheta}}
\newcommand{\iotabf}{\boldsymbol{\iota}}
\newcommand{\kappabf}{\boldsymbol{\kappa}}
\newcommand{\lambdabf}{\boldsymbol{\lambda}}
\newcommand{\mubf}{\boldsymbol{\mu}}
\newcommand{\nubf}{\boldsymbol{\nu}}
\newcommand{\xibf}{\boldsymbol{\xi}}
\newcommand{\pibf}{\boldsymbol{\pi}}
\newcommand{\varpibf}{\boldsymbol{\varpi}}
\newcommand{\rhobf}{\boldsymbol{\rho}}
\newcommand{\varrhobf}{\boldsymbol{\varrho}}
\newcommand{\sigmabf}{\boldsymbol{\sigma}}
\newcommand{\varsigmabf}{\boldsymbol{\varsigma}}
\newcommand{\taubf}{\boldsymbol{\tau}}
\newcommand{\upsilonbf}{\boldsymbol{\upsilon}}
\newcommand{\phibf}{\boldsymbol{\phi}}
\newcommand{\varphibf}{\boldsymbol{\varphi}}
\newcommand{\chibf}{\boldsymbol{\chi}}
\newcommand{\psibf}{\boldsymbol{\psi}}
\newcommand{\omegabf}{\boldsymbol{\omega}}
\newcommand{\Gammabf}{\boldsymbol{\Gamma}}
\newcommand{\Deltabf}{\boldsymbol{\Delta}}
\newcommand{\Thetabf}{\boldsymbol{\Theta}}
\newcommand{\Lambdabf}{\boldsymbol{\Lambda}}
\newcommand{\Xibf}{\boldsymbol{\Xi}}
\newcommand{\Pibf}{\boldsymbol{\Pi}}
\newcommand{\Psibf}{\boldsymbol{\Psi}}
\newcommand{\Sigmabf}{\boldsymbol{\Sigma}}
\newcommand{\Upsilonbf}{\boldsymbol{\Upsilon}}
\newcommand{\Phibf}{\boldsymbol{\Phi}}
\newcommand{\Omegabf}{\boldsymbol{\Omega}}


\newcommand{\sfa}{{\mathsf{a}}}
\newcommand{\sfb}{{\mathsf{b}}}
\newcommand{\sfc}{{\mathsf{c}}}
\newcommand{\sfd}{{\mathsf{d}}}
\newcommand{\sfe}{{\mathsf{e}}}
\newcommand{\sff}{{\mathsf{f}}}
\newcommand{\sfg}{{\mathsf{g}}}
\newcommand{\sfh}{{\mathsf{h}}}
\newcommand{\sfi}{{\mathsf{i}}}
\newcommand{\sfj}{{\mathsf{j}}}
\newcommand{\sfk}{{\mathsf{k}}}
\newcommand{\sfl}{{\mathsf{l}}}
\newcommand{\sfm}{{\mathsf{m}}}
\newcommand{\sfn}{{\mathsf{n}}}
\newcommand{\sfo}{{\mathsf{o}}}
\newcommand{\sfp}{{\mathsf{p}}}
\newcommand{\sfq}{{\mathsf{q}}}
\newcommand{\sfr}{{\mathsf{r}}}
\newcommand{\sfs}{{\mathsf{s}}}
\newcommand{\sft}{{\mathsf{t}}}
\newcommand{\sfu}{{\mathsf{u}}}
\newcommand{\sfv}{{\mathsf{v}}}
\newcommand{\sfw}{{\mathsf{w}}}
\newcommand{\sfx}{{\mathsf{x}}}
\newcommand{\sfy}{{\mathsf{y}}}
\newcommand{\sfz}{{\mathsf{z}}}

\newcommand{\sfA}{{\mathsf{A}}}
\newcommand{\sfB}{{\mathsf{B}}}
\newcommand{\sfC}{{\mathsf{C}}}
\newcommand{\sfD}{{\mathsf{D}}}
\newcommand{\sfE}{{\mathsf{E}}}
\newcommand{\sfF}{{\mathsf{F}}}
\newcommand{\sfG}{{\mathsf{G}}}
\newcommand{\sfH}{{\mathsf{H}}}
\newcommand{\sfI}{{\mathsf{I}}}
\newcommand{\sfJ}{{\mathsf{J}}}
\newcommand{\sfK}{{\mathsf{K}}}
\newcommand{\sfL}{{\mathsf{L}}}
\newcommand{\sfM}{{\mathsf{M}}}
\newcommand{\sfN}{{\mathsf{N}}}
\newcommand{\sfO}{{\mathsf{O}}}
\newcommand{\sfP}{{\mathsf{P}}}
\newcommand{\sfQ}{{\mathsf{Q}}}
\newcommand{\sfR}{{\mathsf{R}}}
\newcommand{\sfS}{{\mathsf{S}}}
\newcommand{\sfT}{{\mathsf{T}}}
\newcommand{\sfU}{{\mathsf{U}}}
\newcommand{\sfV}{{\mathsf{V}}}
\newcommand{\sfW}{{\mathsf{W}}}
\newcommand{\sfX}{{\mathsf{X}}}
\newcommand{\sfY}{{\mathsf{Y}}}
\newcommand{\sfZ}{{\mathsf{Z}}}


\newcommand{\cala}{{\mathcal{a}}}
\newcommand{\calb}{{\mathcal{b}}}
\newcommand{\calc}{{\mathcal{c}}}
\newcommand{\cald}{{\mathcal{d}}}
\newcommand{\cale}{{\mathcal{e}}}
\newcommand{\calf}{{\mathcal{f}}}
\newcommand{\calg}{{\mathcal{g}}}
\newcommand{\calh}{{\mathcal{h}}}
\newcommand{\cali}{{\mathcal{i}}}
\newcommand{\calj}{{\mathcal{j}}}
\newcommand{\calk}{{\mathcal{k}}}
\newcommand{\call}{{\mathcal{l}}}
\newcommand{\calm}{{\mathcal{m}}}
\newcommand{\caln}{{\mathcal{n}}}
\newcommand{\calo}{{\mathcal{o}}}
\newcommand{\calp}{{\mathcal{p}}}
\newcommand{\calq}{{\mathcal{q}}}
\newcommand{\calr}{{\mathcal{r}}}
\newcommand{\cals}{{\mathcal{s}}}
\newcommand{\calt}{{\mathcal{t}}}
\newcommand{\calu}{{\mathcal{u}}}
\newcommand{\calv}{{\mathcal{v}}}
\newcommand{\calw}{{\mathcal{w}}}
\newcommand{\calx}{{\mathcal{x}}}
\newcommand{\caly}{{\mathcal{y}}}
\newcommand{\calz}{{\mathcal{z}}}

\newcommand{\calA}{{\mathcal{A}}}
\newcommand{\calB}{{\mathcal{B}}}
\newcommand{\calC}{{\mathcal{C}}}
\newcommand{\calD}{{\mathcal{D}}}
\newcommand{\calE}{{\mathcal{E}}}
\newcommand{\calF}{{\mathcal{F}}}
\newcommand{\calG}{{\mathcal{G}}}
\newcommand{\calH}{{\mathcal{H}}}
\newcommand{\calI}{{\mathcal{I}}}
\newcommand{\calJ}{{\mathcal{J}}}
\newcommand{\calK}{{\mathcal{K}}}
\newcommand{\calL}{{\mathcal{L}}}
\newcommand{\calM}{{\mathcal{M}}}
\newcommand{\calN}{{\mathcal{N}}}
\newcommand{\calO}{{\mathcal{O}}}
\newcommand{\calP}{{\mathcal{P}}}
\newcommand{\calQ}{{\mathcal{Q}}}
\newcommand{\calR}{{\mathcal{R}}}
\newcommand{\calS}{{\mathcal{S}}}
\newcommand{\calT}{{\mathcal{T}}}
\newcommand{\calU}{{\mathcal{U}}}
\newcommand{\calV}{{\mathcal{V}}}
\newcommand{\calW}{{\mathcal{W}}}
\newcommand{\calX}{{\mathcal{X}}}
\newcommand{\calY}{{\mathcal{Y}}}
\newcommand{\calZ}{{\mathcal{Z}}}

\newcommand{\bbA}{{\mathbb{A}}}
\newcommand{\bbB}{{\mathbb{B}}}
\newcommand{\bbC}{{\mathbb{C}}}
\newcommand{\bbD}{{\mathbb{D}}}
\newcommand{\bbE}{{\mathbb{E}}}
\newcommand{\bbF}{{\mathbb{F}}}
\newcommand{\bbG}{{\mathbb{G}}}
\newcommand{\bbH}{{\mathbb{H}}}
\newcommand{\bbI}{{\mathbb{I}}}
\newcommand{\bbJ}{{\mathbb{J}}}
\newcommand{\bbK}{{\mathbb{K}}}
\newcommand{\bbL}{{\mathbb{L}}}
\newcommand{\bbM}{{\mathbb{M}}}
\newcommand{\bbN}{{\mathbb{N}}}
\newcommand{\bbO}{{\mathbb{O}}}
\newcommand{\bbP}{{\mathbb{P}}}
\newcommand{\bbQ}{{\mathbb{Q}}}
\newcommand{\bbR}{{\mathbb{R}}}
\newcommand{\bbS}{{\mathbb{S}}}
\newcommand{\bbT}{{\mathbb{T}}}
\newcommand{\bbU}{{\mathbb{U}}}
\newcommand{\bbV}{{\mathbb{V}}}
\newcommand{\bbW}{{\mathbb{W}}}
\newcommand{\bbX}{{\mathbb{X}}}
\newcommand{\bbY}{{\mathbb{Y}}}
\newcommand{\bbZ}{{\mathbb{Z}}}

\begin{abstract}
Voice assistants have recently achieved remarkable commercial success. However, the current generation of these devices is typically capable of only reactive interactions. In other words, interactions have to be initiated by the user, which somewhat limits their usability and user experience. We propose, that the next generation of such devices should be able to \emph{proactively} provide the \emph{right} information in the \emph{right} way at the \emph{right} time, without being prompted by the user. However, achieving this is not straightforward,  since there is the danger it could interrupt what the user is doing too much, resulting in it being distracting or even annoying. Furthermore, it could unwittingly, reveal sensitive/private information to third parties. 

In this report, we discuss the challenges of developing proactively initiated interactions, and suggest a framework for when it is appropriate for the device to intervene. To validate our design assumptions, we describe firstly, how we built a functioning prototype and secondly, a user study that was conducted to assess users' reactions and reflections when in the presence of a proactive voice assistant. This pre-print summarises the state, ideas and progress towards a proactive device as of autumn 2018.

\end{abstract}


\begin{teaserfigure}
  \vspace{-0.25cm}
  \includegraphics[width=\textwidth]{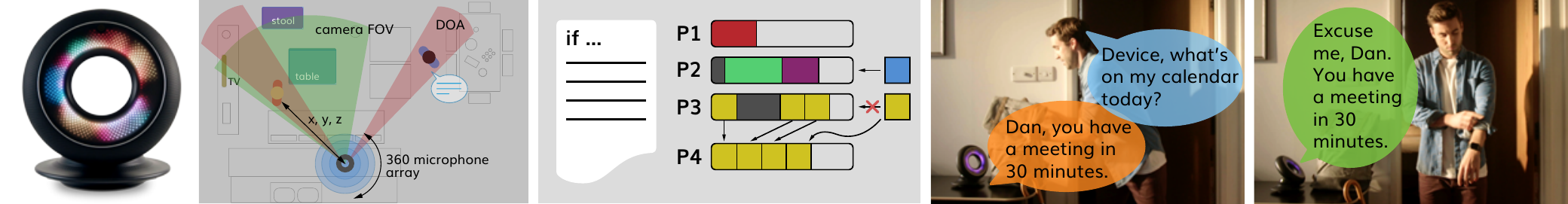}
  \vspace{-0.5cm}
  \caption{We propose that the next generation of voice assistants should be capable of proactive-initiated interactions. We discuss design principles and validate them using a (a) new hardware platform equipped with (b) multi-modal semantic scene understanding and (c) decision making modules. Using the proposed design assumptions, hardware and software, we demonstrate how users can benefit from transforming (d) reactive devices which typically \textcolor{orange}{only respond requests} \textcolor{blue}{initiated by the users}, to the (e) \textcolor{green}{proactive ones} offering the user the \emph{right information} in the \emph{right way} at the \emph{right time} without being asked.}
  \label{fig:teaser}
  \vspace{0.35cm}
\end{teaserfigure}

\maketitle

\section{Introduction}

In 2014, Amazon announced its Echo speaker with Alexa voice-controlled personal digital assistant. Three years later, the smart speakers represented perhaps the fastest growing market in home appliances with more than $30$M devices shipped world-wide \cite{market_analysis}. This trend has been rapidly increasing with recent integrations of smart assistants into various HiFi audio systems~\cite{sonos}, smart TVs~\cite{alexaTv} and cars~\cite{carAssistants}. But what is a voice assistant? What does it currently do? And what else can it \emph{potentially} do? The current generation of voice assistants already performs exceptionally well with basic interactions such as answering knowledge questions (\eg ``What time is it?'' ``Who is the president of the US?''), integrating 3rd party content providers (\eg Spotify, Netflix, \ldots) or controlling various Internet-of-Things (IoT) devices. More advanced products are capable of personalised interactions (\eg ``What is in `\emph{my}' calendar?'' ``Do `\emph{I}' have any unread emails?'') or even short (domain constrained) dialogues \cite{google_duplex}.

However, the current generation of voice assistants is also somewhat limited in the sense that they are \emph{reactive}, \ie they ``only'' \emph{respond} to commands. Moreover, all interactions are initiated by the user using the ``voice trigger'' keyword. Typically, they do not understand their surrounding environments well; they do not understand where they are, what else in the room is, how many people are around or how they interact with each other. Hence, such devices i) may fail in some situations due to the lack of or misinterpreted context (\eg Alexa incident \cite{alexa_incident}) and ii) it is difficult for them to \emph{initiate} non-distracting conversation which significantly limits their capabilities and potential interactions. 

While considerable amount of effort has focused on extending short user-initiated interactions into longer (20 mins or so) multi-domain dialogues \cite{alexa_prize}, we argue, that if voice assistants are to become smarter, they will need to know how to engage in a conversation, and in particular, when to take the initiative to speak. To do so, devices need to be able to proactively assist the users with a range of activities, reminders and day-to-day routines by learning their habits. 

How do designers decide when the device should intervene and what information to use? And how to personalise it for a given person? To begin, we propose that the device should be able to let the user know about important events for the user and more generally in the world. This can be determined and extracted from the user's email accounts/calendar, habits and past digital behaviors. For example the arrival of new email and breaking news about a topic they have shown interest in. It should be announced when it is convenient for the user to attend by assessing their current situation. This could be based on scanning the room for whether the person is alone or with presence of others, what time of day it is, what the person is currently doing and how urgent the information is. At the same time, they should not overload the users with too many verbal updates  or disturb them when they are engaged in another task such as having a conversation with someone else. But how to achieve this so that users find it useful and are comfortable with being interacted with in this manner, while not being annoying or finding it too disruptive or intrusive, is an open question. It is difficult to achieve the right balance due to the inherent ambiguity involved, an expected engagement of user attention and because the consequence of an unwanted distraction is significantly more disturbing and irritating when compared to push-like smartphone notifications, which are less invasive. At the same time, collecting the right kind of data that is not considered invasive of someone's privacy is challenging.

One way to tackle this issue is to analyse the (social) context, using external sensory data collection, that detects certain user "states" such as presence by self, level of busyness, emotional reactions and so on. This approach, however, typically does not scale to data amounts and the diversity required by modern deep reinforcement learning approaches mapping raw audio-visual data directly to decisions \cite{rl} (the reward signal is also too sparse and indirect as we only have an access to weak and noisy proxy such as user emotional states, and potentially very long spans between causes and effects), however, provides sufficient data for designing and studying proactively initiated interactions. An alternative approach is to detect and determine other aspects of the user's context, and their readiness and willingness to be ``spoken'' to by a voice-assisted device in their home. 

In this paper, we describe how we have designed a proactive robot-based voice prototype with the goal of providing the \emph{right information}, in the \emph{right way} at the \emph{right time}, without being prompted by the user. Our approach is to scan the situation using a form of spatial AI and to limit the kinds of proactive interactions to practical day-to-day tasks (\eg weather/traffic/press updates, email/calendar notifications). Our focus is on how to use aspects of the context in relation to a user's privacy. Our approach relies on: (i) semantic scene understanding using spatial AI and multi-modal sensory inputs, (ii) semantic content understanding through prioritising types of interactions, iii) fault-tolerant design of the user experience (UX) and (iv) the design of hardware to draw the user's attention from what they are doing. 

In the first part of the paper,  we describe how we designed our voice assistant prototype to be proactive. It was built using currently available robotic and machine learning technology. To detect context in real time it has been implemented using a novel hardware and software platform equipped with multi-modal sensors. To alert the user to when it is about to speak, the device is programmed to move and light up with a patterned colour display on its body. Then we describe the key elements behind Spatial AI - a method of aggregating relevant statistics across modalities, surrounding space and time. In the remainder of the paper we describe the user study we conducted to evaluate how acceptable, annoying and informative the device was for various conditions, using a living lab experiment. 

\section{Related Work}

The current generation of personal voice assistants is \emph{reactive} in the sense that they ``only'' respond the requests and hence all interactions have to be initiated by the user. 
A typical interaction with a reactive device~\cite{Sarikaya2017TheTB} proceeds as follows: i) in its idle mode, the device is silently ``waiting'' and continuously running a small on-device module whose only purpose is to recognise the ``voice trigger'' keyword (\eg ``Alexa!'', ``Ok Google!'', \ldots); ii) when such a keyword is detected, user provides her request which is streamed to the cloud where this audio input is processed (speech recognition $\rightarrow$ natural language understanding $\rightarrow$ response generation); and iii) the device replies to the user or executes some other interaction (\eg playing a song, setting a timer, \ldots). 

This process is typically repeated from scratch for any other interaction and often even for a simple follow up. More advanced devices are capable to carry the context over a few more exchanges with the user or to offer a ``one more thing''\footnote{Some devices provide ``one more thing'' \emph{during} or at the \emph{end} of an interaction initiated by the \emph{user} (\eg offering a traffic update after being asked for directions), however, this is different from \emph{proactively initiated interaction}.} at the end of an interaction, but not to initiate the interaction itself. What if the smart voice assistants could initiate an interaction or conversation? When and where would it know how to start?

\subsection{Initiating proactivity}
Proactivity for human-robot interactions requires an estimation of \emph{public} and \emph{social} distances~\cite{hall1966hidden} by the robot, so the user is aware of the fact the device exists, is trying to initiate an interaction and the topic of this interaction is also somewhat expected. Various studies~\cite{broadbent2017, yusuke2015, satake2009approach, vaufreydaz2016starting} that have explored how to approach people (for the first time) in the most effective way found that the user's awareness and understanding of the robot's capabilities is crucial to successfully execute pro-activity. 
Alerting people to a new digital event (\eg new text message arrived, breaking news) has been a relatively simple design problem  for \emph{personal} digital assistants embedded on smartphones/laptops/tablets as the public and social distances are better defined from the beginning (physical environment is confined to a mutually known environment, and user understands what to expect from the device). The universal uptake of push notifications on smartphones in the last 10 years has transformed how users are updated of new content; not just new emails or text,  but also likes, new posts and new pictures uploaded. Smith~\etal~\cite{smith2014did} discusses how mobile notification systems affect users and what makes them distracting, and Weber~\etal~\cite{weber2015towards} how to design them such they are less disruptive to the end user. Users can also be in control of how they manage them - choosing to glance, ignore or open the alerting app, and if the continuous stream of notifications becomes too overwhelming, users can switch them off. 

Initiating an interaction from a speech-based robot, however, is quite different. It requires getting someone's  attention and knowing if they are receptive to being interrupted. This involves  determining when the timing is appropriate while also knowing how to best deliver the content verbally (taking into account user engagement at the moment, privacy, efficiency and other contextual information). If the robot butts in at the wrong moment (\eg during an intimate moment) or too often it can be annoying and distracting -- to the point they will abandon using it. However, distraction is not simply a binary matter;  their threshold levels greatly depend on timing (or preceding and current user activity), and that tolerance for their frequency varies between users. This suggests the importance of understanding the contextual setting of the surrounding environment, the value of using personalisation and enabling user adaptation~\cite{smith2014did, mehrotra2016prefminer}. Weber~\etal~\cite{weber2015towards} proposed system of aggregation and distribution of notifications between multiple smart devices (primarily based on the user vicinity to one of them), however the aspects related to \emph{`when'} and \emph{`how'} to notify the user, including corresponding privacy issues in multi-user environments, become paramount. Finding the acceptance threshold may vary from type of notification to type of user. 

Another factor that will become more central in considering proactivity is how the robot is  perceived in terms of its personality, social and emotional intelligence \cite{breazeal2002recognition, breazeal2005effects}. Some people may look forward to their friendly chatty robot telling them things -- akin to having a friendly person at home who is always chatting. Their ability to switch between being proactive and responsive with people needs to be designed to be natural, acceptable and enjoyable. 

The user can choose whether to act upon or ignore a notification appearing on their smartphone or other display. In contrast, voice assistants need to decide when is a good time to notify the user, how many, in what form and in what sequence to present them. One approach to deciding when is an opportune or good time for a virtual assistant to interrupt a user is to analyse conversations in the background \cite{mcmillan2015speech} assuming there is more than one person in the room having a conversation and that the ambient noise (e.g. cooking, TV on) is not too great. If possible this kind of speech recognition could be used to predict when a user might want to run a search on their phone from their conversation, and would require that the speech system is able to detect topical resources from conversation, and be able to perform a level of semantic analysis. 
There are also a number of ethical and privacy concerns with using always on streaming as input for proactive interactions.

The amount of updates a voice assistant might conceivably be in control of is likely to be smaller - at least to begin with - when compared with the number of smartphone push notifications typically received - although this could increase as advertisers and app developers discover ways of attracting 'ears'. The real danger - which is not the case for smartphones - is  the potential to be more disruptive. It only takes one wrongly timed verbal notification to make someone angry. Another challenge is getting the user's attention - especially when they are attending to something else. What kind of signal is required to get someone to listen to the device?  

Few devices to date have been designed to proactively initiated interactions, with some exceptions being social robots like Jibo~\cite{jibo} and Kuri~\cite{kuri} where the presence of the user face (or voice) may trigger some activity conditioned on some auxiliary contextual information (\eg proactive greeting in the morning, invite to play a short game, telling a joke, \etc)\footnote{Note, this report was written in late 2018, when these products were being actively developed. As of now both were cancelled.}. Voice assistants such as Alexa~\cite{alexa} and Google Home~\cite{google_home}, can offer access to personal information (email, calendar) using voice-based authentication, however, in a fully reactive manner, or as a follow up of the ongoing interaction~\cite{Sarikaya2017TheTB}. None offer yet a comprehensive range of reactive and proactive interactions, where the device decides when, what and which information to provide.

Moving from reactive to proactive devices is challenging as it \emph{fundamentally} changes the whole interaction process, requiring advanced cognitive capabilities of devices and to some extent also novel hardware. Consider \eg a ``new email'' proactive reminder to demonstrate the major challenges and key differences from common push-like phone notifications. Smartphones notify the user as soon as the email is received, either by sound, vibration or simply by silently popping the notification up on the screen in case the user does not want to be disturbed, assuming the user will get back to it \emph{whenever it is convenient}. In some sense, a device can (almost) keep flooding the user with more and more notifications as it is the \emph{user who decides (initiates)} what and when is relevant to herself.  This is in sharp contrast to proactive reminders on voice assistants which co-exist with the users in open-world environments (they are not used only when user explicitly controls them). This is a significantly more complex task and the device cannot just ``blindly'' notify the user as soon as an email is received as the user may not even be around. The device therefore has to first possess some comprehension of the surrounding environment and has to identify whether the user is around.  The device has to understand, whether it is convenient to notify the user now as it should not disturb or overload her with too many interactions when she is cognitively engaged (\ie having a conversations or focusing her attention on some other tasks). 

\vspace{-0.2cm}
\subsection{Privacy concerns}
If the device concludes that the user should be notified now, it needs to attract the user's attention before it attempts to deliver the message to give her some time to get prepared and focused. This step is quite different from smartphones, where a subtle buzz or a beep are used. Furthermore, the message needs to be delivered in an appropriate way, based on who is around, \eg some messages may be private or not be appropriate for kids, and therefore should be delivered when the user is alone. For instance, the device should not ask the user whether she is around or who she is as this could quickly become annoying. Instead, it should do this cognitive process in the ``background'' and infer it automatically. Proactive interactions need to be designed in a fault-tolerant manner, taking into account potential AI imperfections. This suggests that they need to err on being conservative, initiating interaction, only when confident the user is willing and ready to listen. 

What happens when they say something wrong? Should they express human-level responses \cite{Hamacher2016BelievingIB} and apologise even (\cf Reeves and Nass~\cite{reeves1996media})? As they  become more proactive, would it be desirable for them to look less like inanimate objects (\eg stationary cylinders) and instead look, animate and behave much more like \emph{robots}?

\subsection{Understanding the local context}
One approach to deciding when is a good time for a robot or smart speaker to alert a user to a new message is to use cues from the local context. Semantic maps have long been considered to be a prerequisite for decision making systems operating in partially observable 3D environments. This problem is known in robotics as Simultaneous Localisation and Mapping (SLAM)~\cite{davison2018spatialai}, while in biological systems as cognitive maps of the environment~\cite{brenden2016arxiv, dolan2013neuron, daw2005nature}. During the past few years, real-time (dense) semantic SLAM has made a significant progress, for instance~\cite{slam++, Hermans2014Dense3S, vineet2015icra, semantic_fusion} showed how to build such maps in real-time using only passive cameras or even learn how to segment previously unseen objects on-the-fly~\cite{semanticpaint, semanticpaintbrush}. 
Bhatti~\etal~\cite{bhatti2016doom} has also shown recently how semantic maps can be used for learning decision making policies for agents operating in dynamic environments. 

Our research  is concerned with what new features and underlying model are needed to enable voice assistants to become smarter by taking the initiative to speak up while avoiding situations where they are perceived to be annoying. 

\section {Methodology}

To build the next generation of voice assistants that have the capability of being both proactive and reactive, our research focuses on the following aspects:
\begin{enumerate}
\item context awareness using spatial AI 
\item semantic content modelling 
\item cueing the user's attention
\end{enumerate}

\subsection{Context awareness using spatial AI}
Spatial AI is a broad term that refers to building representations for decision making of agents operating in spatial domains. As such, in this work it spans scene understanding, speaker and audio event recognition, spoken language processing (including emotion modelling where necessary) and decision making~\cite{davison2018spatialai}
We adopt a multi-modal version of a spatial AI; in-built cameras and a microphone array with computer vision and audio processing algorithms are used to infer a richer picture of what is happening. 

Our approach is to use a combination of multi-modal semantic scene understanding and decision making subsystems to lay the foundation for proactively initiated interactions. Semantic scene understanding accumulates information from multi-modal sensors that provides a single, unified and machine-interpretable overview of the robot's vicinity. A decision making subsystem combines this semantic information about the vicinity using different proactivity levels (see below), user profiles, and meta-data about the past interactions. 
See Sec.~\ref{sec:spatial_ai} for technical details.

\subsection{Semantic content modelling}
The importance of a new message or notification to a user will vary (\eg email with ``Meeting in 10 minutes?'' is likely to be more important than a periodical newsletter). The question this raises is  how does the system decide which is most important and which can wait?  Our approach uses semantic understanding of content of interactions, where messages are hierarchically stacked, ranging from immediate notifications to periodical batch updates. Table \ref{tab:levels} shows our hierarchy of levels of proactivity. Note, these are different from levels of autonomy of a virtual personal assistant \cite{Sarikaya2017TheTB} to independently execute tasks or make decisions on behalf of its owner (\ie automatically decide on things like shopping, booking travel, scheduling meetings, \ldots).

\begin{table}[h]
\centering
 \begin{tabular}{p{\columnwidth}} 
 \hline
\textbf{Level 1 (L1)} -- Comprises push-like (\eg new email) and proactive routines (\eg morning news overview). User recognition enables personalised interactions (\eg greetings, personalised reminders) but no requirement for semantic scene or content understanding. All interactions have equal importance, \ie there is no prioritisation. The users is notified of either using daily batch updates (\eg morning/evening routines) or push-like one-off notifications whenever a new email is received.  All updates have the same priority. \\  
\hline

\textbf{Level 2 (L2)} -- Prioritises messages that are scheduled based on the context. This requires semantic scene and content understanding. A rule that might be used for this is \eg assign higher priorities to breaking news containing keywords (tags) such as ``terrorist attack'' or ``politics''. The use of semantic content understanding enables prioritisation (and ``grouping'') of interactions based on their importance.\\ \hline

\textbf{Level 3 (L3)} -- The highest level, where the voice assistant is capable of life-long learning of user habits to keep refining proactive interactions over time. Proactive interactions are embedded into complex dialogues. It would involve the user and system having more in-depth interactions, for example, helping the user to improve on their well-being. The devices should be able to learn automatically user preferences about content or frequency of updates. \\ 
\hline

 \end{tabular}
 \caption{\label{tab:levels} Proposed hierarchy of proactive interactions.}
 \vspace{-0.5cm}
\end{table}

To begin, we focus on levels 1 and 2, in order to determine if there are any differences between these two levels of proactivity on user acceptance and perceived usefulness. As content, we use practical day-to-day updates, as such interactions transferable across different users. The subset of practical day-to-day interactions we implemented for the study were email, calendar, traffic info, news, IoT lights and TV. See Appendix \ref{app:l_rules} for some examples of L1/L2 rules.

\subsection{Cueing the user's attention}
As part of the move towards creating proactive devices we believe, it is important to consider how to attract the user's attention to when an update/alert is about to be spoken. We decided to incorporate a form of ambient design into the body of our robot voice assistant, through the use of coloured lights, movement and appropriate synchronisation of the UX such the user has enough time to tune into interaction mode with the device. These choices, including hardware and software considerations, are in detail outlined in the following Section~\ref{sec:platform}.

\section{SYSTEM OVERVIEW} \label{sec:platform}
In this section, we describe our platform from hardware, basic user interaction, and software perspectives.
\begin{figure*}[t]
  \includegraphics[width=\textwidth]{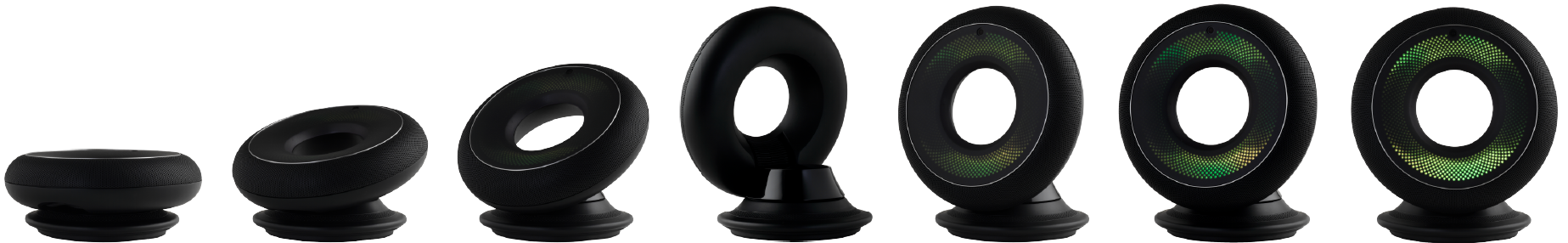}
  \vspace{-0.7cm}
  \caption{Our platform in its ``idle'' (left) and ``up'' (right) states (\cf~Sec.~\ref{sec:user_interface}) with other animations shown in between.}
  \vspace{-0.15cm}
  \label{fig:platform}
\end{figure*}

\subsection{Hardware} \label{ssec:hardware}
\begin{figure}[t]
    \centering
    \includegraphics[width=\linewidth]{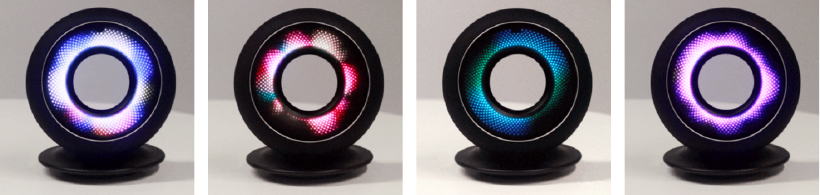}
    \vspace{-0.7cm}
    \caption{LED matrix is able to show arbitrary animations to attract user's attention and express emotions.}
    \vspace{-0.3cm}
    \label{fig:shaders}
\end{figure}

Our device consists of a fixed base and a moving head (\cf~Fig.~\ref{fig:platform} and Fig.~\ref{fig:shaders}). 
It is equipped with two DC motors allowing for continuous 360$^{\circ}$ rotation around its vertical axis and up to 80$^{\circ}$ rotation in direction perpendicular to the vertical and horizontal axes. The front side of the moving head consists of a custom circular-shaped LED matrix with 480 RGB LEDs and covers three speakers. Combination of these components enables the device to attract user's attention, communicate with the user and express various ``emotions'' or mimic persona types~\cite{olly_led} (also see Fig.~\ref{fig:shaders} for a visual example).

\begin{figure}[t]
    \includegraphics[width=\linewidth]{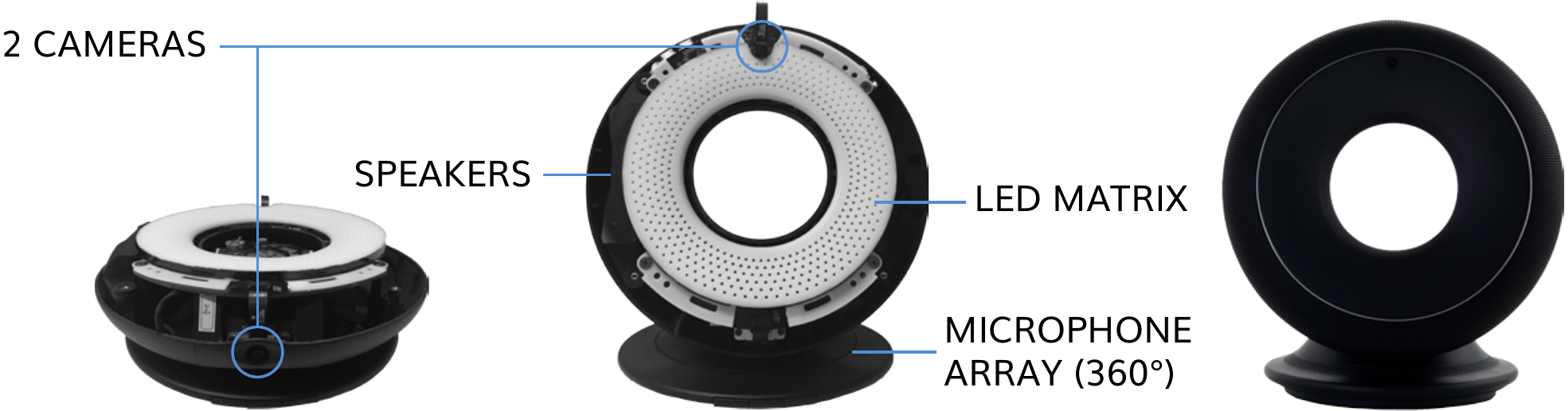}
    \vspace{-0.6cm}
    \caption{Our platform is equipped with two cameras, microphone array, three speakers, LED matrix and two motors.}
    \vspace{-0.1cm}
    \label{fig:hw}
\end{figure}

The moving head contains two 8 mega-pixel RGB wide-angle cameras rotated by 90$^{\circ}$ w.r.t. each other (\cf\figref{fig:hw}) to enable perceiving the surrounding environment under all possible rotations, custom far-field microphone array with $4$ microphones and 6-axis inertial measurement unit (IMU). 
The richer sensory inputs (\ie microphones \emph{and} cameras) allow us not only to process standard audio modality but also combine it with visual data, which significantly extends perception abilities to understand the environment and interactions among the users. 
Thus, it helps to overcome sensor limitations, \eg 360$^{\circ}$ sensing of microphone arrays overcomes limited field-of-view of cameras; visual data may help with source disambiguation in noisy areas, multi-user interactions, \etc 
The IMU is used by motors feedback controllers.
Device uses $6$-core ARM CPU, dual-core GPU with $4$ GB DDR4 RAM, $8$ GB NAND flash memory storage and is equipped with WiFi and Bluetooth modules and runs an embedded Linux OS.

\vspace{-0.15cm}
\subsection{User Interface, Interactions and 3rd Party Services}
\label{sec:user_interface}

\subsubsection{Basic interactions.} Our device has two basic states, called ``\emph{idle}'' and ``\emph{up}'' (\cf\figref{fig:platform}). For the majority of the time, the device is in the ``\emph{idle}'' state waiting for the voice-trigger, which is commonly used in all reactive scenarios. 
However, even in this state, the device detects acoustic events around the device and can rotate around its vertical axis to ``scan'' the 360$^{\circ}$ environment using cameras.  
Scanning is triggered either periodically or using an arbitrary acoustic event (\ie not a hot-keyword; rather sounds corresponding to events such as walking, doors activity, \ldots). The second basic state, ``\emph{up}'', is primarily used when the user interacts with the device or when the device wants to attract user's attention to proactively initiate an interaction with her. 
Our platform supports various transitions between the two and all such animated motions can be combined in arbitrary ways (\cf\figref{fig:platform}).

In a standard reactive scenario, the device is in its ``\emph{idle}'' mode listening for a voice trigger. 
Once this is provided by the user, the device ``\emph{wakes up}'' and rotates so that it faces the user (using \emph{Spatial AI} described below) to establish ``\emph{eye contact}'' with the user. 
Next, the device is ready to process user's request as any other voice assistant would. However, it is also able to express emotions using a combination of animated movement and LED matrix (\cf~\figref{fig:shaders}), \eg when the device does not understand the user's request. Such basic movements and interactions create so called ``\emph{presets}'' that can be arbitrarily combined to create more advanced (non-verbal) interaction capabilities.

\subsubsection{Unboxing scenario.} 
To support personalised interactions (potentially with multiple users), our device has to learn how to recognise the user(s) first.
This happens during the so-called ``\emph{unboxing}'' or ``\emph{learn me}'' interaction, which can be triggered by the user. 
The device instructs the user to move to various locations in the room to collect multiple views of her face that are used to extract 128 dimensional facial embeddings trained using the triplet loss \cite{omkar2015bmvc}, and similar 128 dimensional speaker embeddings obtained from neural network trained using a teacher-student approach~\cite{ng2018}.

\subsubsection{3rd party services.} 
Our device supports numerous services for non-personalised interactions such as weather forecast, headlines and traffic updates as well as personalised interactions such as calendar or email. Our platform also integrates smart TVs, lights, Nest or Sonos as examples of IoT devices.

\begin{figure*}[t]
\centering
  \includegraphics[width=0.98\textwidth]{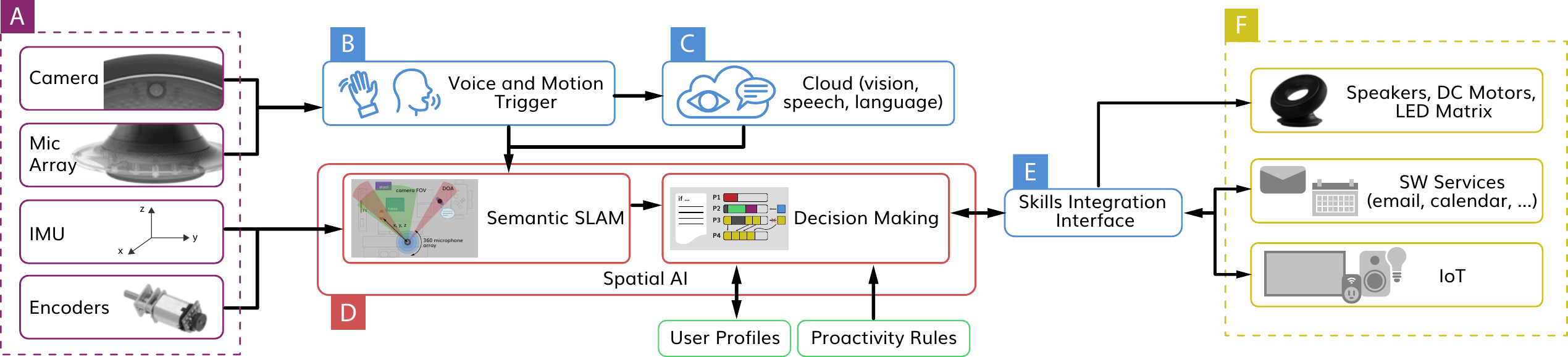}
    \vspace{-0.2cm}
    \caption{Overview of our software pipeline (refer to text for details).}
    \vspace{-0.3cm}
  \label{fig:sw}
\end{figure*}

\subsection{Software} 
Our software pipeline consists of several subsystems (\figref{fig:sw}). 
First, we capture multi-modal data using camera, microphones, IMU and encoders (\figref{fig:sw} A). 
Audio-visual data are passed to on-device continuously running voice and motion trigger, whose only purpose is to recognise the hot keyword or detect the motion (\figref{fig:sw} B). 
When presence of a user is detected and the device has some update ready, or a voice-trigger is spotted, the device wakes up and faces the user \cite{swietojanski2020mmdoa}, starting to process at the same time the audio-visual data. This, depending on the compute requirements can happen either on device or in the cloud (\figref{fig:sw} C) to run more computationally expensive models for computer vision, speech recognition and natural language processing (\eg to map user query to actionable outcome \cite{liu2019benchmarking}). When necessary, some additional attributes like user's emotional state \cite{beard2018multi} or acoustic events \cite{shi2019teacher} may be also estimated.
This output is then sent back to the device and combined with data from IMU and encoders in the \emph{Spatial AI} module (\figref{fig:sw} D), which builds a \emph{semantic map} of the environment and is responsible for all \emph{decision making}. 
This subsystem is supported by a database of user profiles (user preferences, history of past interactions, \etc) and proactivity rules. 
The Spatial AI block is connected with a Skills integration interface (\figref{fig:sw} E), which executes actions (play sound, rotate robot, control LED), two-way communication with software services (email, calendar, \ldots) and IoT devices (\figref{fig:sw} F). 

\section{SPATIAL AI} \label{sec:spatial_ai}
At the heart of our device lies the \emph{Spatial AI} module~\cite{davison2018spatialai}, a combination of multi-modal semantic scene understanding and decision making subsystems which lays the foundation for proactively initiated interactions:

1. Semantic scene understanding accumulates information from multi-modal sensors available on our platform and provides a single, unified and machine-interpretable overview of the robot's vicinity. Note, that having a representation capable of accumulating statistics across time is also beneficial to enable lifelong (incremental) learning of users' habits, however, this is beyond the scope of the paper (L3 devices).

2. Decision making subsystem combines semantic information about vicinity with proactivity rules, user profiles, meta-data about the past interactions, and is responsible for prioritising and scheduling interactions.

\begin{figure*}[t]
    \centering
    \includegraphics[width=\textwidth]{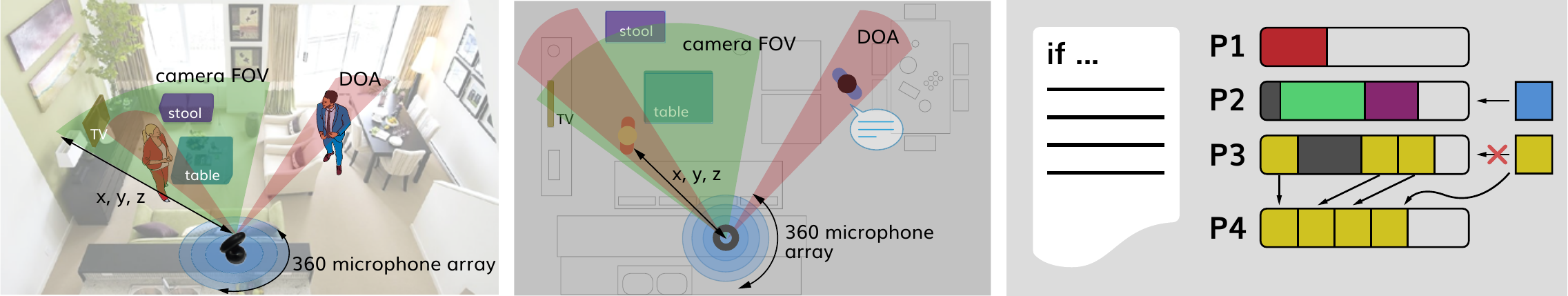}
    \vspace{-0.7cm}
    \caption{Spatial AI. Semantic scene understanding accumulates information from multi-modal sensors (left) and provide a single, unified and machine-interpretable overview of the robot's vicinity (middle) which is used by decision making (right).}
    \vspace{-0.3cm}
    \label{fig:spatial_ai}
\end{figure*}

\vspace{-0.15cm}
\subsection{Semantic Scene Understanding}
We draw inspiration from \cite{bhatti2016doom} and opt for ``top-down'' views (re-projections) of semantic maps (\cf\figref{fig:spatial_ai}). To this end, we estimate localization of the robot with respect to the environment and in parallel detect important stationary (\eg TV, sofa, \ldots) and dynamically moving objects (users). 
In order to update a semantic map from robot's first-person view at each frame, we accumulate such semantic information by projecting it onto a common 2D map, essentially a ``floor-plan'' with encoded positions of the robot and objects. 

\subsubsection{Model.} 
We use a multi-modal tracking-by-detection paradigm with probabilistic data association formulated as a Markov Random Field (MRF) \cite{Zhang08globaldata, pgm}. 
Let $\calX = \{\bfx_i\}$ denote a set of observations corresponding to detection responses $\bfx_i = \{x_i, t_i, a_i, d_i, l_i\}$ where $x_i$ is the position, $t_i$ the time step, $a_i$ appearance and audio features, $d_i$ the detection score and $l_i$ the semantic label. 
A trajectory $T_k$ is defined as an ordered sequence of observations $T_k = \{\bfx_{k1}, \bfx_{k2}, \ldots, \bfx_{k_l}\}$, where $\bfx_{k_l} \in \calX$. 
The goal of the global data association is to maximize the posterior probability of trajectories $\calT = \{T_k\}$ given the set of observations $\calX$
\begin{equation}
    p(\calT|\calX) = p(\calT) \prod_i p(\bfx_i|\calT)        \label{eq:data_association}
\end{equation}
The likelihood function $p(\bfx_i | \calT)$ of the observation $\bfx_i$ is defined by Bernoulli distribution which models the cases of being a true detection as well as false alarm
\begin{equation}
p(\bfx_i |\calT) \approx    
\begin{cases}
    \beta_i,        & \text{if} \quad \exists\, T_k \in \calT \;\land \; i \in T_k \\
    1 - \beta_i,    & \text{otherwise}
\end{cases}
\end{equation}
The prior over trajectories decomposes into the product of unary and pairwise terms
\begin{equation}
    p(\calT) \approx \prod_{T \in \calT} \psi(T) \prod_{T, T' \in \calT} [T \cap T' = \emptyset]
\end{equation}
where the pairwise term ensures the trajectories are disjoint. The unary term is given by
\begin{equation}
    \psi(\calT) = \psi_{en}(\bfx_{k_1}) \, \psi_{ex}(\bfx_{k_l}) \prod_{i=1}^{l-1} \psi_{li}(\bfx_{k_i}, \bfx_{k_{i+1}})
\end{equation}
where $\psi_{en}$, $\psi_{ex}$ and $\psi_{li}$ encode likelihood of entering a trajectory, exiting a trajectory and linking temporally adjacent observations within a trajectory. Note, that our representation could also naturally accommodate dense(r) representations (semantic segmentation, material prediction, \ldots) and dense 3D reconstruction if needed, as it has been shown in \cite{bhatti2016doom}.

\subsubsection{Inference.} 
Taking a negative logarithm of \eqref{eq:data_association} turns the maximization into an equivalent energy minimization problem which can be mapped into a min-cost flow network and efficiently solved using an online min-cost solver with bounded memory and computation \cite{Zhang08globaldata, Lenz2015ICCV}. 
We periodically re-run this inference step in an asynchronous thread.

\subsubsection{Appearance features.} 
We use similar association features to Lenz \etal~\cite{Lenz2015ICCV}, \ie an LAB colour histogram, patch similarity, bounding-box overlap, bounding-box size, location and class label similarity (\cf \cite{Lenz2015ICCV} supp.). 
In order to detect the bounding boxes, we exploit prior knowledge about the scene. 
The stationary objects (TV, sofa, \ldots) are detected using the YOLO object detector~\cite{redmon2016yolo9000} running as an asynchronous service in a cloud. 
The predicted bounding-boxes are directly fed into the MRF. 
However, this would result in too large latency for user detection, tracking and recognition (they are not stationary). 
Therefore, we run a second, lightweight, dlib frontal face detector \cite{dlib} on a device GPU which (re)-initializes the fast DSST trackers~\cite{dsst, staple} running in asynchronous threads to achieve interactive framerates. 
Such outputs are used directly (\eg to maintain an ``eye contact'' with the user within the camera frustum) and as inputs into the MRF. 
Note, that we could have used a single model suitable for embedded devices such as MobileNet \cite{mobilenets}, however, this is rather an implementation detail beyond the scope of the paper.

\subsubsection{Audio features.} 
For acoustic event detection, we use log mel filterbank features extracted from a raw audio signal followed by a convolutional neural network producing per-class posterior probabilities \cite{audioFilterbank, shi2019teacher}. 
To take into account co-occuring audio events, we notify the spatial model about each acoustic event that surpasses the expected threshold. 
Additionally, we estimate direction of arrival (DOA) $\theta_s$ for each of the detected sounds $'s'$ using a set of DOA estimates from the raw signal (as many as detected acoustic events at each given time step), which are then mapped to $x,y,z$ coordinates\footnote{In far-field, one cannot easily estimate the distance between sound source and microphone array, thus we assume constant radius when mapping from polar to Cartesian coordinates.}. This process can leverage an additional semantic information from vision stream, as shown in \cite{swietojanski2020mmdoa}. 
The most likely pairs \{\texttt{acoustic\_event}, $\theta_s$\} for co-occurring events are estimated in the spatial model using visual data.

\subsubsection{User recognition for personalised interactions.} 
Whenever we detect a face or a spoken acoustic event, we extract an embedding vector and associate it with a particular object trajectory $T_k$. 
At each observation, the embedding vector is classified as a known or unknown user (\ie open-set recognition) using standard feature thresholding and discriminating w.r.t. other known users and mean (background) models. %
The confidence scores are accumulated across time to avoid per-frame independent decisions and ``flickering'' predictions. 

\vspace{-0.15cm}
\subsection{Decision Making}
Learning a decision making agent is non-trivial due to the lack of training data and sparsity of the reward signal.
Additionally, our primary goal is to validate our design assumptions. 
Hence, we use a manually designed first-order logic decision rules. 
This makes the system flexible enough (we can quickly modify interactions) and at the same time remains easily interpretable (easy to understand failures).

\vspace{-0.15cm}
\subsubsection{Interactive rules.} 
Let $\calR = \{\bfr_i\}$ be a set of interactive rules defined by tuples $\bfr_i = \{p_i, e_i, t^s_i, t^c_i, o_i\}$, where $p_i$ is the priority, $e_i$ the time span since previous interaction, $t^s_i$ a triggering service (\eg received email), $t^c_i$ is the triggering configuration (\eg interact if user is the only person around) and $o_i$ is the set of output actions (LED, speaker, \ldots). 
Note that multiple rules can be combined together by using them as triggering service $t^s_i$ (\eg ``weather update'' can be appended to ``calendar reminder''). 
We run an asynchronous thread periodically checking all active rules and their associated trigger events. 
Note that multiple interactions might be triggered at the same time or before the current one finishes. 
Hence, all triggered interactions are pushed into the scheduler to ensure the user is not overloaded. 
This does not prevent reactive interactions initiated by the user using a voice trigger; such interactions are simply pushed into the scheduler with the highest priority reserved for the reactive mode, \ie immediate responses to requests initiated by the keyword phrase.

\vspace{-0.15cm}
\subsubsection{Scheduling.} 
We use a multilevel feedback queue scheduler \cite{operating_systems}, which groups interactions into $P$ queues.
We use linked lists implementation to support iterating over jobs and job removal from the middle of the queue.
Each queue is assigned a priority and has its own scheduling algorithm; we use first-in-first-out scheduling. 
This ensures that an interaction is executed when all the queues with higher priority have been completed. 
In contrast to a multilevel queue, jobs can move between the queues which prevents starvation of lower priority tasks, and ``jobs recombination'' can transform multiple tasks in the same queue into a batch (\eg $N$ email newsletter updates $\rightarrow$ single update ``the user has $N$ emails'') and either push it back to the same queue, promote to a higher or demote to a lower priority queue (\cf~\figref{fig:spatial_ai}, right).

\vspace{-0.15cm}
\subsubsection{Accumulating meta-data.}
For each trajectory $T_k$, we maintain a fixed size queue of last $N$ interactions, time and priority of their execution.
Such statistics are essential for multi-level scheduling described above.
Thus we propagate it to the user profiles, from which, it can be retrieved to help with situations when a user \eg leaves the room for a few minutes.

\subsection{Computational Efficiency and Scalability}
In addition to on-device processing resources, the current spoken language understanding stack (speech recognition + natural language understanding) runs in the cloud with final latency of around $200-300$ms as measured from the point when the user has finished her query. 

Part of the computer vision stack (object detection, facial embedding extraction) which runs in cloud uses Nvidia \mbox{Titan X} and average processing time takes around $150-200$ ms. 
However, it should be noted this latency influences only the first ``user detection'', as then we interleave detection with fast on-device object trackers running at $20$fps.

Clearly, the amount of high-end hardware required to run our prototype is relatively high, however, it i) is possible to replace many computationally expensive parts of our pipeline by their lightweight alternatives suitable for embedded devices such as MobileNet \cite{mobilenets}; ii) not every single device needs to have dedicated hardware, it should rather be shared by multiple devices which could efficiently use minibatching to optimize the cost.

\section{USER STUDY} \label{sec:user_study}

We conducted a user study in a living lab, set up to determine how people would react to our proactive robot that is implemented using our spatial AI model. The study was designed primarily to investigate how varying the amount and type of digital content with respect to L1 and L2 interactions impacted on how users found the different kinds of updates to be useful, distracting or even annoying. We also investigated the extent to which different contexts affected the user's perceptions and what was considered an acceptable update frequency in differing contexts (by varying the scenario and frequency of updates) and what was the effect of being alone or in the presence of someone else. Another variable we were interested in was how to get someone's attention when they are involved in another task. Would a dynamic cue manifested in the device's head orienting towards the user and its LED pattern appearing be able to draw their attention without it being annoying? To measure people's perceptions and reactions, we observed them during the study when subjected  to different kinds of proactive interactions initiated by the device and then interviewed them afterwards. Two scenarios were set up: one during a routine period of the day and the other during a non-routine lazy part of the day. This enabled us to explore whether level of busyness affected acceptance and perceived usefulness of the proactive interruptions.
\\

\noindent \textbf{L1 versus L2 levels.}
The content proactively spoken by the device was for practical day-to-day routines. These were: email, calendar notifications, headlines, weather and traffic updates.
Half the sessions were run as L1 interactions - where the device is able to recognise the user, but where no spatial modeling is taking place. The updates are one-off. The other half were run as L2 interactions, where the system uses spatial modelling to detect where the participants are and what they were doing. This provided us with a base to decide when to pro-actively intervene.

For either mode, we used deterministically pre-defined digital content (user's mailbox, calendar, news \ldots) to ensure reproducibility for all participants. The news content was synthesised by creating fake news (e.g. \emph{Donald Trump has resigned}). The relative importance was determined by its assumed level of interest. For example, the news headline \emph{Donald Trump has resigned} was assigned to be more important than \emph{Tesco starts selling cars}.

For both L1 and L2 levels, the user was presented with the same number of messages in a session. For L1, however, their arrival was random, and our device notified the user as soon as they were received. L2, on the other side, prioritised and batched the arriving messages. Batching was done based on importance, but also privacy considerations (\ie personal message even if important, should not be read in case user is not alone).
\\

The context was varied for the study. \textbf{Condition 1} was designed to simulate relatively short repetitive parts of a day, (e.g. morning routines). \textbf{Condition 2} was designed to simulate longer interactions (e.g. lazy afternoon or weekends). As such Condition 1 was designed to take 20 minutes, while Condition 2 was longer and took 40 minutes. The number and types of messages announced in each mode is reported in Table~\ref{tab:conditions}. Example email messages for the case where the device was expected to preserve privacy are shown in the Appendix~\ref{app:personal}.\\

\begin{table}[h]
\centering
 \begin{tabular}{l|c|c|c} 
  & Email & Calendar & Other updates  \\ \hline
 Condition 1 & 6 & 4 & 2  \\ \hline
 Condition 2 & 16 & 6 & 4 \\ \hline
 \end{tabular}
 \caption{\label{tab:conditions} Types and the number of messages communicated to participants in each condition.}
 \vspace{-0.5cm}
\end{table}

\noindent \textbf{Static versus animated device.}
To test the importance of attracting people's attention before speaking an update, the device was programmed to work in two modes (i) scanning the room in its ``idle'' mode, and (ii) using an animated motion with lights appearing on the display. This process is shown in \figref{fig:platform}, where on the left device is in its sleeping ``idle'' state, smoothly transitioning to the ``wake'' interaction mode on the right. The duration of this transition was configurable, but in our experiments took around 2 seconds.\\

\noindent \textbf{Alone or in presence of other person.} 
The study was designed so that for half the time participants sat with someone else and the other by themselves. Participants did not know each other prior to the session. The device operated either in L1 or L2 modes (for the whole session).

\subsection{Participants and Protocol}
The study took place in London, UK. 20 participants (10 females and 10 males), aged between 18 and 36, were recruited and asked to come three times over a period of two weeks on different days to the living lab. Each participant was paid 15 pounds per hour. 

Two participants, who did not know each other prior to the meeting, were brought into the lab that was furnished like a living room. The device was set up in the corner of the room, while the participants were sitting on a couch. The rest of the room was furnished in a usual way, with chairs, coffee table, shelves, wardrobe and TV. 

The participants were familiarised with the experiment and device with an instruction brief (attached in Appendix~\ref{app:instructions}). The instruction shortly stated the purpose of the research, the plan for the following two sessions as well as the top level summary of what the device could and could not do in the study. Participants were also instructed about the quizzes, cognitive tasks, an exit questionnaire and the importance in participating in the other two sessions in different days. They were then asked if they had any follow up questions, which were then answered verbally.

The first participant was asked to engage in a primary task - which depending on the session was either watching a short movie, or reading a story. Apart from this cognitively engaging task, participants were encouraged to interact with each other as they would normally do if they were sharing home environment.

For the condition 1, the participants were asked to imagine they had been away for a while (\ie morning after waking up), so the device has accumulated lot of content for them, including new emails, remainder and press news. Condition 1 tested how the device should approach the notification process, i.e. in the order messages arrived or in a batched manner taking into account priorities of each notification, as well as privacy concerns. 
In L2 the participant was distracted sporadically, using contextual cues to try to prevent them from becoming overloaded. This session lasted about 20 minutes. After the first session, the participants were given a short break of 15 minutes. 

The second session took 40 minutes and focused on Condition 2 content. For this session the participants were told to imagine that they are enjoying a lazy weekend afternoon in order to better understand how their preferences about frequency and types of updates changes with the amount of time spend with the device. The other difference was the device did not have anything upfront to announce, but rather tried to keep up with the arriving content. In L1 mode, these were being announced immediately as they arrived while in L2 the device tried to prioritise, batch and preserve privacy.

The rationale for having these two different scenarios was to investigate degree of distraction someone is happy to accept under various situations (idle, chatting with others, cognitively engaged on some task) and how important it is to attract user's attention prior to the interaction.

The idle and conversation aspects were embedded by design (users had some time alone with a device, as well as were encouraged to speak with each other). For the part where participants should focus on some cognitively engaging task, they were asked to answer some quiz like questions. For condition 1 these were short movie clips (i.e. 5 minutes long) and the corresponding quiz participants were asked to solve (based on the content of the video). Condition 2 involved a 10 minute long reading exercise and the related answer quiz. These were to assess to what degree the device's announcements affect the participants' ability to focus under different operation modes (L1/L2). As shown in the instruction form (see Appendix~\ref{app:instructions}), the participants were instructed to give answers only when they managed to learn the answer (no need for guessing).

The order of conditions were counterbalanced as well as the order of situations using the Latin square design. During each session, two researchers observed the two participants' behavior and their interaction patterns with the device. Half way through each session one participant was asked to leave the room. The reason for this is that we also wanted to test how participants would react when just by themselves. Would they feel more comfortable? How would being by oneself differ from being with someone else when the robot proactively spoke to them? How would their approach change when device shared private messages?

At the end of the 2nd session in each day, the participants filled in the survey (see see Appendix~\ref{app:answers}) that asked them about their experience of a device notification system, and several short questions about the content of notifications comparing their experiences of whether L1 or L2 approaches were perceived to be more accurate in delivering content.

\subsection {Findings}
The observations made during the sessions and the answers from the survey revealed that overall most participants appreciated the benefits of having proactively initiated interactions. In general, they liked the appearance of the device, and were not too concerned that it needed to scan the room in idle state. P4  commented, ``Enjoyed the proactivity of the robot, no need to check your phone for updates'', P6: noted  ``concept of the machine knowing what updates to give and when''. The least impressed participant (P15) agreed that ``proactive updates are useful in terms of interaction''. Participants also commented on liking the variety of content (about meetings, calendar events, traffic and weather). Some said they would have appreciated having some control over pausing or stopping them. For example,  P15: said there should be the option to ``pause updates if there are too many... or simply opt-out''. P5 wanted to be able to ``ask whether the user wants details or simply skip to speed up going through too many accumulated news'. P6: would have liked to be able to ``customize / filter updates'' while P2 wanted to be able to ``request the device to repeat the last update''.

Many of the updates read out what turned out to be long texts (email, news). Participants wanted these to be shorter summaries as it was hard for them to pay attention for these lengths of time. Other participants complained about not being able to control the volume of the updates (P10) and the tempo of updates.

\subsubsection{(i) Differences between L1 and L2.} 
There were not many differences in the participants' comments for the low-frequency content, only a few participants noticed the ranking was not good enough for L1 settings (\ie lack of priorities in this mode resulted in switching contexts between different types of messages).
However, this differed dramatically for high-frequency content, where participants showed a strong preference towards L2. In particular, the participants appreciated the assigned priorities and updates given in batches for the L2 settings, for example,  P15 noted: ``Updates are good. Importance ranking is good'', while P8 and P10 both liked batch updates of news.

This difference was even more pronounced for the L1 setting, which participants found very invasive and disturbing. For example, P14: stressed how they ``couldn't focus on my task'', P6 said ``I found it invasive, couldn't concentrate on the tasks'', P13 pointed out ``there were too many updates, barely had time to think'', P3, P7 and P17 all said that there was a lot of information which was quite distracting. 

Similarly, participants did not like the way the updates were just announced with no apparent reason behind them. P11 for example, said they ``found it very annoying as it just threw information at me in a random order'',  while P1, P2, P15 and P16 said they would prefer if the important updates had been read first for batched updates. Others would have liked to have more control over the length of email and news updates that were read out. 

Participants in general liked that when the device was set to L2, it postponed all personal updates until they were alone (note, in L1 device did not use spatial AI module, thus did not track if there is more than one person in the room and as such could leak some content of the private correspondence. More on this in the next sections.). Participants also pointed out that updates should not be given when they were talking with others, which point towards another important aspect of spatial understanding of the environment.

\subsubsection{(ii) Stationary versus animated device.} 
Most of the participants liked the animated device, for example, P2 liked that the device faced me when giving an update'', P11 also liked ``the way it moved before giving any updates ''P16, thought that it acted eye-contact , giving ``a good idea of when the device will speak'' and P7 and P9 thought it was good that the ``device wakes up before giving an update''. Conversely, participants complained when the device provided an update without attracting their attention in this way -- having gotten used to it. P11, for example, noted how ``I didn't like how it gave no warning that it was about to update us so sometimes I missed the first part of the update'',  while P3 was annoyed as  ``I wasn't prepared for the updates'', and P9 also said ``I didn't like that it did not wake up before giving an update''.

\subsubsection{(iii) Naive vs privacy-preserving updates.}
Almost all participants immediately found personal updates in the presence of other people annoying, for example P11 said ``Found it very annoying as it told me private things when someone else was in the room.''. See examples in Table~\ref{tab:emails} in the Appendix~\ref{app:personal}. 
On the other hand, participants much preferred the L2 type interactions with personal updates given when they were alone. P7 for example, said ``It was great when the device said: 'Looks like we're finally alone, so I can update you about personal matters'''. In fact, the importance of privacy was the most widely discussed topic that was raised. One participant (P9) became aware of how it could be problematic, ``the door was open when updating me, meaning others could listen in easily''.  To deal with the privacy concern, several participants suggested that the device should request permission before giving out personal information (P1, P10, P12, P15).

\section{DISCUSSION and CONCLUSIONS}

In this paper, we have proposed that \emph{proactively initiated interactions} will be one of the defining factors of the next generation of voice assistants, identified design principles for such devices, and since not all interactions are equally complex, defined their classification levels.
In addition to that, we have described an end-to-end prototype comprising of novel hardware and software, which we used to conduct a live-lab user study to validate our design assumptions.

One of the key assumptions our user study confirms is, that the privacy truly matters to the users\footnote{The study was conducted in London, UK. This might differ in other parts of the world.}. 
In fact, it represents the key challenge for \emph{proactively initiated interactions} as majority of \emph{important} updates are typically quite personal.
As such, it not only introduces new demands on the hardware and software stack to ensure the updates are provided only when sensitive information cannot be accidentally leaked to the 3rd parties, but also calls for strong legislation and data protection (most voice assistants are used in home environments).
From this perspective, it is highly positive to see efforts of some governments putting such legislation in place (most notably GDPR); and to see that many companies go even beyond and consider data privacy to be a \emph{human right}. 
We would like to stress, that while such legislation and compliance comes with certain (development) costs, it is absolutely critical; not just for the end users perspective, but also for the actual progress of smart voice assistants as it ``protects'' the \emph{developers} (suppresses fear that the technology they are contributing could be easily misused).

On a hardware side, our prototype device is equipped with the necessary sensors and features to implement proposed pro-activity levels (both perception-- and expression-- wise). The proposed Spatial AI model is generic enough for fusing various sources of sensory and high level information required to understand the immediate environment and to accordingly initiate interactions. However, if such device were to be deployed beyond the lab environment one would need to carefully communicate when, how and where the device is using the information it has access to, both in real-time when making ad-hoc decisions (\ie it is clear for the user when cameras/mics are on/off and what information exactly the device is looking for in each stream) but also beyond that (\ie is the information stored for further processing / models re-estimation? If so, where - on the device or in the cloud? Who has access to it? May it be manually inspected or annotated at any point? Can the user conveniently reply and manage stored episodes? \etc). Some of these issues are being addressed by legislation (like the aforementioned GDPR) but much remain to the terms and conditions and often implementation details. 

In the experiments carried in this work, the participants did not raise device-related privacy concerns, but it does not mean they would not have any if the device was at their place. Likewise, we did not answer if they would be happy with the additional privacy trade offs required to provide good experience of proactive interactions (cameras / mics on) when compared to the current reactive devices (only mics on). In either case one would require configurable mechanisms and related functionality to back off to more limited capabilities if the user wishes to (temporarily) disable any of the modalities.

Finally, we would like to stress that while proactively initiated day-to-day interactions (email, calendar, press, \ldots) exhibit promising potential and demonstrated benefits to the users, we are only at the very beginning.
It might be very tempting to start promising interactions proactively improving users' well-being, and in general, imitating user's \emph{best friend} (proactive suggestions to ``go to a therapy'', ``have a glass of wine'' or ``told joke to improve user's mood''). 
However, we need to keep in mind that such interactions are much less transferable across different users, often depend on user's personality, current mood and in general require much better understanding of (cultural, social, \ldots) context.
While AI has been making great progress, with its current state, we are nowhere near devices that could support such interactions. 
Thus, we need to select interactions which we try to transform into proactively initiated very carefully.

\subsection{Acknowledgements}
In no particular order, we would also like to thank X. Chen, M. Zhou, A. Ye, J. Zhao, P. Tesh, J. Grant, A. Khan, J-C Passepont, G. Groszko, M. Shewakramani and T. Wierzchowiecki for their contributions during various stages of this project.

\bibliographystyle{ACM-Reference-Format}
\bibliography{sample-bibliography}

\clearpage

\begin{appendices}

\section{Examples of decision making rules} \label{app:l_rules}

\subsection{L1 rules:}

\begin{itemize}
\item Personalised greetings:
\begin{algorithmic}
\If {user\_detected\_1st\_time \& user\_recognised}
  \State message $\gets$ Hi \$User
\ElsIf {user\_detected\_1st\_time \& !user\_recognised}
  \State messsage $\gets$ Hi
\EndIf
\end{algorithmic}
\item Weather:
\begin{algorithmic}
\If {user\_detected\_1st\_time\_a\_day}
  \State update about weather
\EndIf
\end{algorithmic}
\item IoT lights:
\begin{algorithmic}
\If {user\_detected \& time $\leq$ 9am}
  \State turn lights on
\EndIf
\end{algorithmic}
\item Calendar:
\begin{algorithmic}
\If {event $\leq$ 2 hours}
  \State remainder\_priority $\gets$ high
\EndIf
\end{algorithmic}
\begin{algorithmic}
\If {event\_same\_day}
  \State remainder\_priority $\gets$ medium
\Else
  \State remainder\_priority $\gets$ low
\EndIf
\end{algorithmic}
\item Other (email, news, etc):
\begin{algorithmic}
\If {new\_event}
  \State put into scheduler (push-like)
\EndIf
\end{algorithmic}
\end{itemize}

\subsection{L2 rules (services):}

\begin{itemize}
\item Email:
\begin{algorithmic}
\If { whitelisted \{family, boss, friends\} email }
    \State $importance$ $\gets$ $high$
\ElsIf {spam or newsletters}
    \State $importance$ $\gets$ $low$
\EndIf
\end{algorithmic}
\item News:
\begin{algorithmic}
\If {  contains \{terrorist, politics\} }
    \State $priority$ $\gets$ $high$
\EndIf
\end{algorithmic}
\end{itemize}
\subsection{L2 rules (scheduling / meta-rules):}

\begin{itemize}
\item 
\begin{algorithmic}
\If {personal\_update \{email, calendar\}}
  \State postpone until user is alone
\EndIf
\end{algorithmic}
\item 
\begin{algorithmic}
\If {event\_type exists in Queues}
  \If {event\_importance is high}
     \State combine into a single one-by-one interaction
   \Else
      \State combine into batched interaction
   \EndIf
\EndIf
\end{algorithmic}
\item 
\begin{algorithmic}
\If {news \& weather}
  \State first update news
\EndIf
\end{algorithmic}
\item 
\begin{algorithmic}
\If {news \& calendar}
  \State first update important news then unimportant calendar
\EndIf
\end{algorithmic}
\item 
\begin{algorithmic}
\If {news \& email}
  \State first update email
\EndIf
\end{algorithmic}
\item 
\begin{algorithmic}
\If {calendar \& email}
  \State first update important calendar
\EndIf
\end{algorithmic}
\item 
\begin{algorithmic}
\If {time\_last\_update - time\_elapsed s < $T_{min}$}
  \State schedule interaction in ($T_{min}$ - time\_elapsed)
\EndIf
\end{algorithmic}
Note, that $T_{min}$ is a constant specific for each interaction type (and priority)
\end{itemize}

\section {Examples of personal messages} \label{app:personal}

Here we show examples of "personal" notifications, for which device should preserve privacy (read them only when user is alone).

\begin{table}[h]
\centering \small
 \begin{tabular}{p{\columnwidth}} 
 \hline
You have a new email Lloyds bank. It says: Hi [user], You still have not paid back your debt and you have only 500 pounds at your account. Therefore your credit card will be blocked. \\
\hline
You have a new email from Your Boss. It says: Hi [user], I'm very much unhappy with your performance and hence decided to put you on performance improvement plan. You have 6 months to prove your value, otherwise you will be terminated. Let's have a chat about it later today. \\
\hline
You have a new email from Thames Water. It says: Hi [user], please give us meter readings by end next week \\
\hline
You have a new email from Pedro. It says: Hey [user], pub at 6? \\
\hline
[user], I'm supposed to remind you a date with Anna tonight. \\
\hline 
You have a new email from Jeniffer. It says: Hey [user], dinner at my place at 7? \\
\hline
[user], I'm supposed to remind you to fire an intern next week. \\
\hline
You have a new email from Jeniffer. It says: Stop stalking me!!!!! Next time I'm gonna call police you bastard!!! \\
\hline
 \end{tabular}
 \caption{\label{tab:emails} Examples of the private messages used during living lab sessions. The [user] tag was used to personalise device interactions with each participant.}
 \vspace{-0.5cm}
\end{table}


\section{Participant forms}

\subsection {Entry instructions} \label{app:instructions}
\includepdf[scale=0.8]{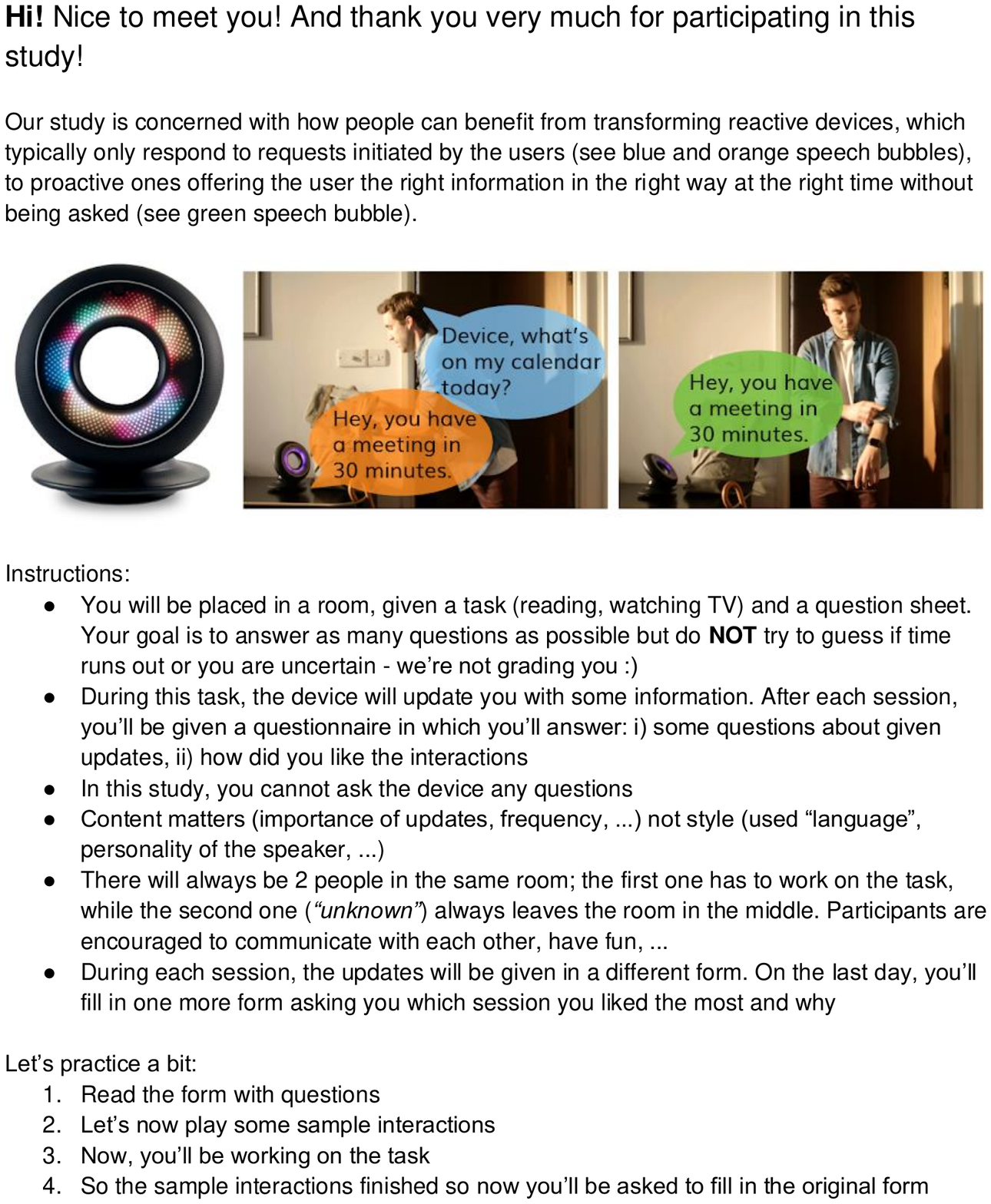}

\subsection {Answer sheet} \label{app:answers}
\includepdf[scale=0.8]{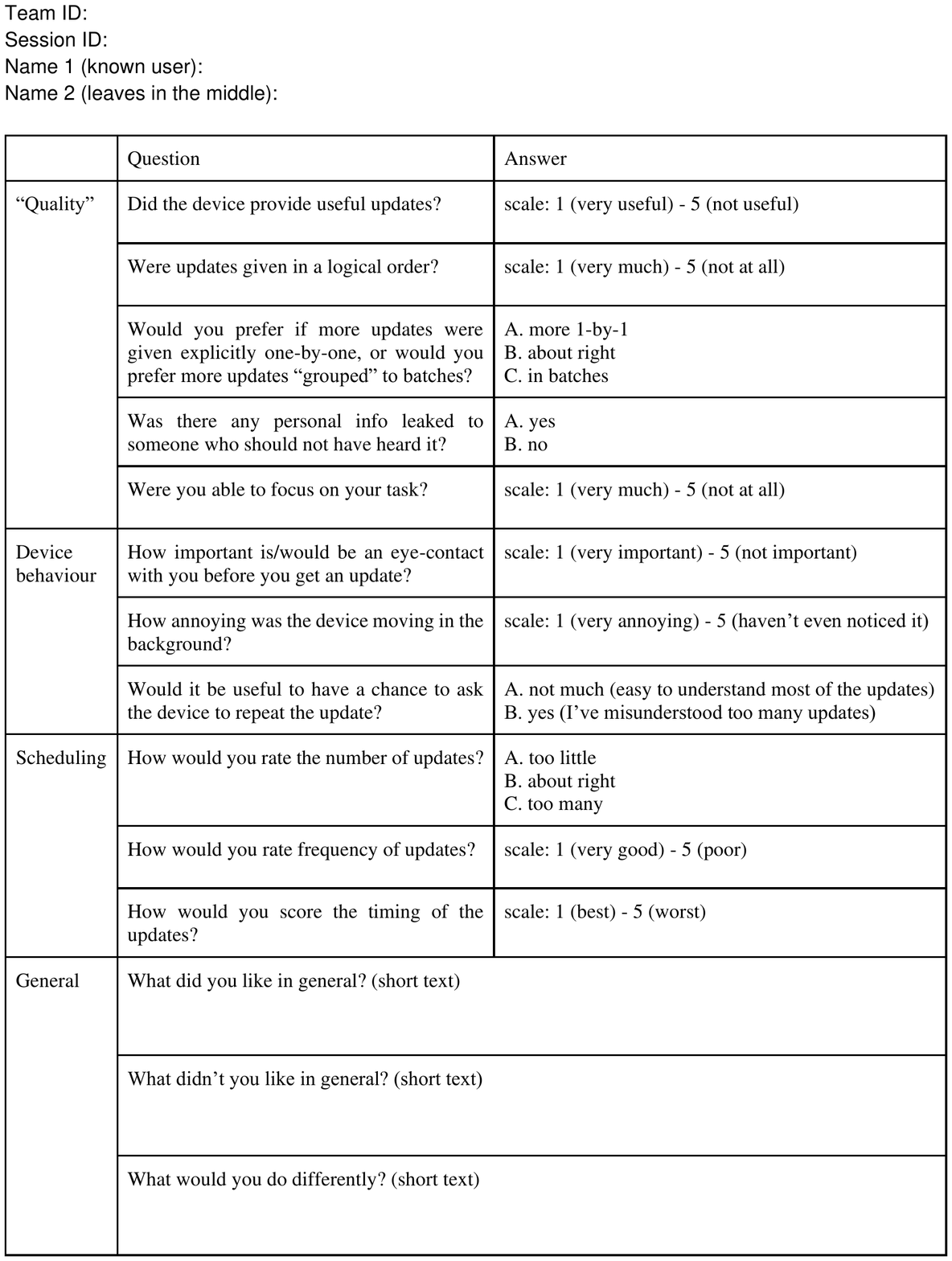}

\subsection {Exit questionnaire} \label{app:exit}


\section*{Supplementary material}

# C  PARTICIPANT FORMS
**Entry instructions**

**Hi!** Nice to meet you! And thank you very much for participating in this study!

Our study is concerned with how people can benefit from transforming reactive devices, which typically only respond to requests initiated by the users (see blue and orange speech bubbles), to proactive ones offering the user the right information in the right way at the right time without being asked (see green speech bubble).

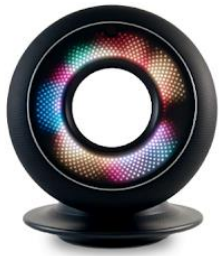
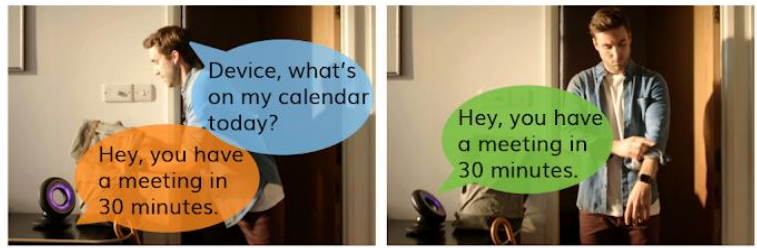

Instructions:
- You will be placed in a room, given a task (reading, watching TV) and a question sheet. Your goal is to answer as many questions as possible but do **NOT** try to guess if time runs out or you are uncertain - we're not grading you :)
- During this task, the device will update you with some information. After each session, you'll be given a questionnaire in which you'll answer: i) some questions about given updates, ii) how did you like the interactions
- In this study, you cannot ask the device any questions
- Content matters (importance of updates, frequency, ...) not style (used "language", personality of the speaker, ...)
- There will always be 2 people in the same room; the first one has to work on the task, while the second one (*"unknown"*) always leaves the room in the middle. Participants are encouraged to communicate with each other, have fun, ...
- During each session, the updates will be given in a different form. On the last day, you'll fill in one more form asking you which session you liked the most and why

Let's practice a bit:
1. Read the form with questions
2. Let's now play some sample interactions
3. Now, you'll be working on the task
4. So the sample interactions finished so now you'll be asked to fill in the original form

# Answer sheet

Team ID:
Session ID:
Name 1 (known user):
Name 2 (leaves in the middle):

|  | Question | Answer |
|---|---|---|
| "Quality" | Did the device provide useful updates? | scale: 1 (very useful) - 5 (not useful) |
|  | Were updates given in a logical order? | scale: 1 (very much) - 5 (not at all) |
|  | Would you prefer if more updates were given explicitly one-by-one, or would you prefer more updates "grouped" to batches? | A. more 1-by-1<br>B. about right<br>C. in batches |
|  | Was there any personal info leaked to someone who should not have heard it? | A. yes<br>B. no |
|  | Were you able to focus on your task? | scale: 1 (very much) - 5 (not at all) |
| Device behaviour | How important is/would be an eye-contact with you before you get an update? | scale: 1 (very important) - 5 (not important) |
|  | How annoying was the device moving in the background? | scale: 1 (very annoying) - 5 (haven't even noticed it) |
|  | Would it be useful to have a chance to ask the device to repeat the update? | A. not much (easy to understand most of the updates)<br>B. yes (I've misunderstood too many updates) |
| Scheduling | How would you rate the number of updates? | A. too little<br>B. about right<br>C. too many |
|  | How would you rate frequency of updates? | scale: 1 (very good) - 5 (poor) |
|  | How would you score the timing of the updates? | scale: 1 (best) - 5 (worst) |
| General | What did you like in general? (short text) | |
|  | What didn't you like in general? (short text) | |
|  | What would you do differently? (short text) | |

# Exit questionnaire

Team ID:
Session IDs:
Name:

| Which day / session did you like most? (you can describe the device behavior by words if you don't remember the day/session) |
|---|
|  |

| What did you like in general? (short text) |
|---|
|  |

| What didn't you like in general? (short text) |
|---|
|  |

| What would you do differently? (short text) |
|---|
|  |

| Bonus - would you by device with such capabilities? How much would you pay for it? |
|---|
|  |



\end{appendices}

\end{document}